\documentclass[prd,nofootinbib,superscriptaddress,twocolumn,floatfix]{revtex4}

\usepackage{times} %
\usepackage{pdfsync}%

\usepackage{xcolor} %
\definecolor{lightgray}{gray}{0.9}

\usepackage{amssymb,amsfonts,amsmath,paralist,xspace} %

\usepackage{longtable}%
\usepackage[normalem]{ulem} %
\usepackage{relsize,color,xcolor} %
\usepackage{booktabs} 

\usepackage{units} 

\usepackage{grffile} %
\usepackage{graphicx} %
\graphicspath{{./figures/}} 

\IfFileExists{srcltx.sty}{\usepackage[active]{srcltx}}

\usepackage{etoolbox} %
\preto\tabular{\setcounter{magicrownumbers}{0}}%
\newcounter{magicrownumbers}%
 \newcommand{\be}
{\begin{equation}} \newcommand{\ee} {\end{equation}}   
\newcommand{\eq}[1]{\begin{equation} #1 \end{equation}}
\def\be{\begin{eqnarray}} \def\ee{\end{eqnarray}} 
 \renewcommand{\S}{\mathcal{S}} %
\newcommand{\dm}{{\textsc{dm}}} 
\newcommand{\xmm}{\textit{XMM-Newton}\xspace}
\newcommand{\fov}{\mathrm{fov}} 
 
\renewcommand{\S}{\mathcal{S}} 

\newcommand{\nfw}{\mathrm{NFW}} 
\begin{document}

\title{Testing the origin of $\sim$3.55 keV line in individual galaxy clusters observed with XMM-Newton}

\author{Dmytro~Iakubovskyi$^1$, Esra Bulbul$^2$, Adam R.~Foster$^2$, Denys Savchenko$^1$ and Valentyna Sadova$^3$\\ 
{\small $^1$ Bogolyubov Institute of Theoretical Physics, Metrologichna Str. 14-b, 03680, Kyiv, Ukraine \\
 $^2$ Harvard-Smithsonian Center for Astrophysics, 60 Garden Street, Cambridge, MA 02138, USA \\
 $^3$ Taras Shevchenko National University of Kyiv, Physics Department, Glushkova ave. 2, Kyiv, Ukraine}
 }

\begin{abstract}
If the unidentified emission line at $\sim$3.55~keV previously found in spectra of nearby galaxies and 
galaxy clusters is due to radiatively decaying dark matter, one should detect the signal of comparable strength 
from many cosmic objects of different nature. By studying existing dark matter distributions in galaxy clusters 
we identified top-19 of them observed by \xmm\ X-ray cosmic mission, and analyzed the data for the presence of 
the new line. In 8 of them, we identified $>$~2$\sigma$ positive line-like residuals 
with average position 3.52$\pm$0.08~keV in the emitter's frame. Their observed properties 
are unlikely to be explained by statistical fluctuations or astrophysical emission lines; observed 
line position in M31 and Galactic Center makes an additional argument against general-type systematics. 
Being interpreted as decaying dark matter line, the new detections correspond to radiative decay
lifetime $\tau_\dm \approx (3.5-6)\times 10^{27}$~s consistent with previous detections.
\end{abstract}

\maketitle

The origin of missing mass in cosmic objects ranging from dwarf galaxies to galaxy clusters, large-scale 
structure and the observable part of our Universe, remains unknown. Assuming Newtonian/Einsteinian gravity 
and dynamics to be valid one has to introduce new type of matter -- the \emph{dark matter} -- 
presumably in form of new elementary particles beyond the Standard Model of particle physics. 
Non-zero interaction strength of dark matter particles with Standard Model particles may lead to \emph{2-body 
radiative decay} of the dark matter. Because present-day velocities of dark matter particles in haloes 
should be highly non-relativistic, such process would produce the narrow \emph{dark matter decay line}. 
This motivates extensive ongoing studies of such lines in spectra of cosmic objects with established 
dark matter contribution.

The most intriguing dark matter decay line candidate reported so far is the $\sim$3.55~keV line 
detected in central part of the Perseus galaxy cluster and different combinations of 
galaxy clusters~\citep{Bulbul:14a}, Andromeda galaxy 
and Perseus galaxy cluster outskirts~\citep{Boyarsky:14a}, Galactic Center~\citep{Boyarsky:14b},
Perseus, Coma, and Ophiuchus galaxy clusters~\cite{Urban:14}
\footnote{Based on observed $\sim 60-100$~eV shift between line positions, 
\protect\cite{Urban:14} found improbable that their detections in Coma and Ophiuchus
are of the same origin as in Perseus.}, see also~\cite{Iakubovskyi:14}. 
Although decaying dark matter can naturally explain basic properties of the line in these objects
-- correct scaling of the line signal with object redshift and projected dark matter mass density --
several alternative hypotheses invoking standard or anomalously enhanced astrophysical line 
emission~\citep{Jeltema:14a,Jeltema:14b,Carlson:14}, decay of excited dark 
matter states~\cite{Cline:14c}, 
annihilating dark matter~\cite{Dudas:14,Frandsen:14}, 
dark matter decaying into axion-like particles with further conversion to photons in magnetic 
field~\cite{Alvarez:14}
have been proposed thereafter, see~\cite{Iakubovskyi:14} and references therein.

To further check the decaying dark matter origin of the $\sim$3.55~keV line, we identified 
cosmic targets having the largest expected decaying dark matter signal. We used 
public observations of European Photon Imaging Camera (EPIC)~\citep{Turner:00,Strueder:01} 
on-board \xmm\ X-ray observatory~\citep{Jansen:01} -- the most sensitive existing instrument to search 
narrow faint X-ray lines~\citep{Boyarsky:06f}.

\textit{Object selection}.
For distant objects, the dark matter decay flux [$\unit{in\ photons\ cm^{-2}\ s^{-1}}$] is
\eq{F = \frac{\S_\dm \Omega_\fov}{4\pi m_\dm \tau_\dm},\label{eq:taudm}}
where $\S_\dm \equiv \int \rho_\dm(l) dl$ is the dark matter column density along the line of sight,
$\Omega_\fov \ll 1$ is the Field-of-View of the instrument, 
$m_\dm$ -- mass of the dark matter particle, $\tau_\dm$ -- radiative dark matter decay lifetime. 

We define signal-to-noise ratio as
\eq{SNR = \frac{N_{line}}{\Delta N_{back}},} where $N_{line} \propto \S_\dm t_{obs}$ 
is the number of counts expected from decaying dark matter during observation time $t_{obs}$ 
 and $\Delta N_{back}$ is the uncertainty of background counts.
Neglecting systematical errors and assuming Gaussian statistics, we obtained 
$\Delta N_{back} = \sqrt{B t_{obs}}$, where $B$ is background count rate (in cts/s) 
measured in 3.4-3.65~keV in object's rest frame. 
So the signal-to-noise ratio is proportional to
\eq{SNR \propto S_\dm \sqrt{\frac{t_{obs}}{B}}.}

Because of strong dependence of $SNR$ from $S_\dm$, we first identified objects with the largest 
column density in \xmm/EPIC Field-of-View. In this paper, we concentrated on galaxy clusters -- 
the objects having the largest column densities inside the characheristic radius of dark 
matter haloes~\citep{Boyarsky:09b,Boyarsky:09c}. For each of these objects, 
we calculated dark matter column 
densities $S_{obj}$ inside the innermost 14' radius circle $R_{14}$ 
(roughly corresponding to the \xmm/EPIC Field-of-View) 
using dark matter distributions available in~\cite{Boyarsky:09b}, see 
Appendix~\ref{app:dm-distributions},
and broadened the obtained column density ranges by 0.15~dex to account typical residual uncertainties 
in dark matter distributions~\cite{Boyarsky:09b}.
We then calculated the foreground column density $\S_{MW}$ from Milky Way halo
by using the newest dark matter distributions compiled in~\cite{Boyarsky:14b}. Because in most of objects 
\xmm/EPIC energy resolution is comparable to the energy split between the expected dark matter decay signals 
from object and the Milky Way halo, we multiplied the latter by a correction factor:
\eq{\S_\dm = \S_{obj} + \S_{MW}\times \exp\left(-\frac{z_{obj}^2 E_{line}^2}{2\sigma_{\text{instr}}^2}\right),} 
where $z_{obj}$ is the object redshift, $E_{line} \approx 3.55$~keV is the line position, 
$\sigma_{\text{instr}}$ is the Gaussian dispersion width corresponding
to energy resolution of the instrument\footnote{For Gaussian line, the energy resolution is characterized by 
full width at half-maximum (FWHM) of the line which is $2\sigma\sqrt{2\log(2)}\approx 2.35\sigma_{\text{instr}}$.}. 
According to Fig.~5.24 of~\cite{Iakubovskyi:13}, for \xmm/EPIC imaging spectrometers $\sigma_{\text{instr}} \approx 60$~eV 
at the energies of our interest.

We identified 20 galaxy clusters with the largest $S_\dm$,  
one of them -- Abell~539 -- was not observed by \xmm. The basic properties of the 
remaining 19 objects are summarized in Table~\ref{tab:clusters-main}.

\textit{Data reduction}.
For objects of our interest, we first downloaded all public observation data files for MOS~\citep{Turner:00} 
and PN~\citep{Strueder:01} cameras of \xmm\ X-ray observatory~\citep{Jansen:01}, and
processed them using \emph{Extended Sources Analysis Software} (ESAS)
package~\citep{Kuntz:08} 
 publicly available as part of \xmm\ Science
Analysis System (SAS) v.14.0.0. Time intervals affected by highly variable background component -- soft proton
flares~\cite{Kuntz:08} -- are filtered using ESAS 
scripts \texttt{mos-filter} and \texttt{pn-filter}. 
We used the standard filters and cuts provided by ESAS software. 
We excluded bright point sources detected with the standard SAS procedure \texttt{edetect\_chain},
extracted source spectra and produced response matrices 
inside the 14' radius circle around the NASA Extragalactic Database (NED) source position using ESAS procedures
\texttt{mos-spectra} and \texttt{pn-spectra}. Background spectra were prepared by ESAS scripts 
\texttt{mos\_back} and \texttt{pn\_back}. 
For PN camera, we additionally corrected the obtained spectra for 
out-of-time events. 
 Finally, for each object we combined spectra and response files 
from MOS and PN cameras using \texttt{addspec} FTOOL procedure similar to~\cite{Boyarsky:14a,Boyarsky:14b}, 
and grouped the obtained spectra by 60~eV per energy bin to make the bins 
roughly statistically independent.

\textit{Spectral modeling}. For each object, we modeled separately its combined MOS and PN
spectra in \texttt{Xspec} spectral package with the sum 
of non-thermal (\texttt{powerlaw}) and thermal (line-free \texttt{apec}) continuum components, 
and several narrow \texttt{zgaussian} 
lines of astrophysical origin and the new line absorbed with \texttt{phabs} model and folded with response files.
The \texttt{powerlaw} index $\Gamma = 1.41$ and normalization 11.6~$\unit{photons\ 
cm^{-2}\ s^{-1}\ sr^{-1}\ keV^{-1}}$
at 1~keV are fixed to best-fit values of~\cite{DeLuca:03}.
We chose the modeled energy range 2.1-6.0~keV avoiding strong emission lines. 
To account residual \emph{line-like} calibration uncertainties 
we added 0.4\% (MOS) and 0.25\% (PN) systematic errors in quadratures using \texttt{Xspec} 
parameter \texttt{systematic}, according to Sec.~5.3.5 of~\cite{Iakubovskyi:13}.
The absorption hydrogen column density was fixed at weighted Leiden-Argentine-Bonn survey~\cite{Kalberla:05} 
value obtained through n$_{\text H}$ tool of the NASA High Energy Astrophysics Science Archive Research Center
(HEASARC). 
The redshifts of  
\texttt{apec} and \texttt{zgaussians} were fixed at values from NASA Extragalactic Database (NED). 
The new line position is allowed to vary in 3.35-3.70~keV. To calculate the 
contribution of other astrophysical lines in this region, we used the bright `reference' S~XV line complex at 
$\sim$2.63~keV detected in majority of our combined spectra,
see Appendix~\ref{app:lines-modeling} for details.

Fit quality, plasma temperatures, maximal expected fluxes from K~XVIII line complex at 3.51~keV, 
new line positions and normalizations, and increase of $\chi^2$ statistics due to 
the new line are summarized in Table~\ref{tab:clusters-model}. 
If no line is detected at 1$\sigma$ level, we put 2$\sigma$ upper bounds instead. 

\begin{table*}
 \centering
 \caption{Galaxy clusters observed by \xmm\ ranged by expected significance of decaying dark matter signal from 
 their central parts.}
 \label{tab:clusters-main}
 \vspace*{1ex}
 \begin{tabular}{lcclcc}
  \hline
Object & redshift & $\S_\dm$, M/pc$^2$ & ~~~~~~~~~~~~~~~\xmm ObsID & MOS/PN exposure, ks~~~~ & MOS/PN $SNR$, arb. units \\ 
(1)    &    (2)   &         (3)        & ~~~~~~~~~~~~~~~~~~~~~~~~~~~~~~(4)        &      (5)  & (6)       \\
\hline
Virgo      & 0.0036 & 624-1338 & 0108260201, 0110930701, 0210270101 & 193.2 / 62.9 & 3.1-6.7 / 2.3-5.0  \\
Centaurus  & 0.0114 & 818-1721 & 0046340101, 0406200101             & 348.2 / 123.7 & 2.9-6.0 / 2.2-4.6 \\
Abell~85   & 0.0551 & 130-777 & 0065140101, 0723802101, 0723802201  & 401.6 / 136.7 & 0.6-3.8 / 0.5-2.7 \\
Abell~478  & 0.0881 & 83-969 & 0109880101                           & 111.8 / 43.2 & 0.3-3.5 / 0.2-2.7 \\
Abell~2199 & 0.0302 & 123-826 & 0008030201, 0008030301, 0008030601, & 254.4 / 97.5 & 0.5-3.4 / 0.4-2.6 \\
           &        &         & 0723801101, 0723801201              &              &  \\
Abell~496  & 0.0329 & 124-772 & 0135120201, 0506260301, 0506260401  & 250.6 / 81.0 & 0.5-3.1 / 0.4-2.2 \\
2A0335+096 & 0.0363 & 75-608 & 0109870101, 0109870201, 0147800201   & 225.2 / 89.0 & 0.4-2.9 / 0.3-2.4 \\
Abell~1060 & 0.0126 & 451-1420 & 0206230101                         & 66.7 / 24.7 & 0.9-2.8 / 0.8-2.4 \\
Abell~3266 & 0.0589 & 385-768 & 0105260701, 0105260801, 0105260901, & 179.8 / 63.9 & 1.4-2.8 / 1.1-2.2 \\
           &        &         & 0105261001, 0105261101, 0105262101, &              &  \\
           &        &         & 0105262201, 0105262501              &              &  \\
Abell~S805 & 0.0139 & 286-660 & 0405550401, 0694610101              & 92.2 / 12.5 & 1.2-2.7 / 0.5-1.1 \\
Coma       & 0.0231 & 191-1193 & 0124711401, 0153750101, 0300530101, & 343.8 / 122.0 & 0.4-2.6 / 0.3-2.1 \\
           &        &         & 0300530301, 0300530401, 0300530501, &              &  \\
           &        &         & 0300530601, 0300530701              &              &  \\
Abell~S239 & 0.0635 & 256-553 & 0501110201                          & 81.0 / 28.1 & 0.9-1.9 / 0.6-1.3 \\
Abell~2142 & 0.0909 & 88-573 & 0674560201                           & 104.8 / 40.9 & 0.3-1.7 / 0.2-1.3 \\
Abell~2319 & 0.0557 & 359-716 & 0302150101, 0302150201, 0600040101  & 159.7 / 60.7 & 0.8-1.5 / 0.6-1.2 \\
Abell~1795 & 0.0625 & 83-589 & 0097820101                           & 71.7 / 23.2 & 0.2-1.5 / 0.2-1.1 \\
Abell~209  & 0.2060 & 67-500 & 0084230301                           & 33.5 / 11.2 & 0.2-1.4 / 0.1-1.0 \\
Perseus    & 0.0179 & 418-871 & 0085110101, 0305780101              & 316.4 / 44.2 & 0.7-1.4 / 0.3-0.7 \\
PKS0745-191 & 0.1028 & 59-458 & 0105870101                          & 31.9 / 5.2 & 0.1-0.8 / 0.1-0.5 \\
Triangulum & 0.0510 & 379-757 & 0093620101                          & 18.6 / --- & 0.3-0.6 / --- \\
\hline
 \end{tabular}
\end{table*}

\begin{table*}
 \centering
 \caption{Model parameters of MOS/PN combined spectra of galaxy clusters listed in the previous Table. 
 Line positions are given in cluster's rest frame. Column (4) shows our estimate on maximal K~XVIII line flux 
 at 3.51~keV using prominent S~XVI line complex at 2.63~keV, see Appendix~\ref{app:lines-modeling} for details.
 Errors on line position and flux are given at 1$\sigma$ level for 2 d.o.f. calculated using $\Delta\chi^2 = 2.3$. 
 Line fluxes are in 
 $\unit{10^{-6} ph\ cm^{-2}\ s^{-1}}$, abundances are in Solar values given by~\protect\cite{Anders:89}.
 The new line is detected at $> 2$~$\sigma$ (corresponding to $\Delta\chi^2 > 6.2$) in 8 objects (marked in bold),
 confirming previous detections in Perseus~\protect\cite{Bulbul:14a,Urban:14} and Coma~\protect\cite{Urban:14}.
 }
 \label{tab:clusters-model}
 \vspace*{1ex}

 \begin{tabular}{clcccccc}
  \hline
No & Object  & $\chi^2$/d.o.f. & $T_{e,line}$, keV &  
max K flux at 3.51~keV~~~~~~ & New line position, keV & New line flux & $\Delta \chi^2_{line}$ \\ 
& (1)    &    (2)   &         (3)        &       (4)        &      (5) & (6) & (7)  \\
\hline
1 & Virgo      & 68.9/48 / 60.0/43 & 1.4 / 1.4 & $< 0.9$ / $< 0.7$ & 3.38$^{+0.05}_{-0.05}$ / --- & 4.2$^{+3.1}_{-3.3}$ / $< 9.3$ & 3.8 / 0.4 \\
2 & Centaurus  & 64.1/47 / 64.3/46 & 2.2 / 2.2 & $< 5.8$ / $< 5.6$ & 3.51$^{+0.12}_{-0.19}$ / --- & 25.2$^{+19.4}_{-24.2}$ / $< 15.6$ & 2.9 / 0.1 \\
3 & \textbf{Abell~85} & 37.5/49 / 61.9/46 & 2.2 / 3.4 & $< 0.7$ / $< 0.9$ & \textbf{3.44$^{+0.06}_{-0.05}$} / --- & \textbf{6.3$^{+3.9}_{-3.6}$} / $< 4.2$ & 7.0 / 0.0 \\
4 & Abell~478  & 54.1/47 / 48.9/49 & 2.2 / 1.4 & $< 0.4$ / $< 0.5$ & --- / --- & $< 20.4$ / $< 13.6$ & 0.1 / 0.4 \\
5 & \textbf{Abell~2199} & 46.8/47 / 70.6/52 & 2.7 / 2.7 & $< 1.8$ / $< 1.6$ & \textbf{3.41$^{+0.04}_{-0.04}$} / --- & \textbf{10.1$^{+5.1}_{-4.8}$} / $< 10.0$ & 10.2 / 0.1 \\
6 & \textbf{Abell~496}  & 36.3/45 / 66.7/48 & 3.4 / 2.2 & $< 1.3$ / $< 0.5$ & \textbf{3.55$^{+0.06}_{-0.09}$} / \textbf{3.45$^{+0.04}_{-0.03}$} & \textbf{7.5$^{+6.1}_{-4.4}$} / \textbf{16.8$^{+5.9}_{-6.4}$} & 6.2 / 18.8 \\
7 & 2A0335+096 & 70.9/49 / 65.3/49 & 2.2 / 2.7 & $< 1.5$ / $< 1.3$ & --- / --- & $< 15.5$ / $< 10.7$ & 0.5 / 1.6 \\
8 & Abell~1060 & 67.4/48 / 63.0/50 & 2.2 / 2.7 & $< 2.9$ / $< 1.8$ & --- / --- & $< 27.1$ / $< 21.2$ & 0.2 / 0.0 \\
9 & \textbf{Abell~3266} & 40.7/47 / 67.2/50 & 1.7 / 1.7 & $< 0.3$ / $< 0.3$ & 3.64$^{+0.05}_{-0.08}$ / \textbf{3.53$^{+0.04}_{-0.06}$} & 6.5$^{+4.3}_{-5.3}$ / \textbf{8.7$^{+5.1}_{-4.5}$} & 3.9 / 8.0 \\
10 & \textbf{Abell~S805} & 49.0/45 / 33.4/26 & 1.7 / 1.4 & $< 0.2$ / $< 0.3$ & --- / \textbf{3.63$^{+0.05}_{-0.06}$} & $< 8.7$ / \textbf{17.1$^{+9.3}_{-7.4}$} & 0.3 / 10.8 \\
11 & \textbf{Coma}       & 41.2/37 / 54.7/48 & 2.2 / 4.3 & $< 1.9$ / $< 2.0$ & \textbf{3.49$^{+0.04}_{-0.05}$} / 3.41$^{+0.11}_{-0.10}$ & \textbf{23.7$^{+10.7}_{-9.0}$} / 14.8$^{+9.2}_{-9.6}$ & 16.6 / 3.5 \\
12 & Abell~S239 & 56.7/48 / 60.8/52 & 1.4 / 1.7 & $< 0.1$ / $< 0.2$ & --- / --- & $< 12.3$ / $< 13.6$ & 0.3 / 0.5 \\
13 & Abell~2142 & 63.9/50 / 56.9/50 & 1.4 / 1.4 & $< 0.3$ / $< 0.3$ & --- / --- & $< 9.8$ / $< 17.4$ & 0.0 / 0.8 \\
14 & \textbf{Abell~2319} & 49.4/47 / 61.6/51 & 1.4 / 2.2 & $< 0.4$ / $< 1.4$ & \textbf{3.59$^{+0.05}_{-0.06}$} / 3.53$^{+0.11}_{-0.21}$ & \textbf{18.6$^{+10.7}_{-7.4}$} / 10.5$^{+12.6}_{-10.2}$ & 13.9 / 2.4 \\
15 & Abell~1795 & 61.5/51 / 64.6/50 & 1.7 / 1.7 & $< 0.3$ / $< 0.5$ & --- / ---  & $< 12.4$ / $< 16.5$ & 0.7 / 0.0 \\
16 & Abell~209  & 62.3/50 / 68.0/48 & 1.4 / 1.4 & $< 0.5$ / $< 0.2$ & --- / ---  & $< 17.4$ / $< 9.4$ & 0.6 / 0.0 \\
17 & \textbf{Perseus}    & 69.6/48 / 81.2/47 & 2.7 / 2.7 & $< 4.5$ / $< 6.1$ & \textbf{3.58$^{+0.05}_{-0.08}$} / --- & \textbf{25.2$^{+12.5}_{-12.6}$} / $< 70.4$ & 9.8 / 0.7 \\
18 & PKS0745-191 & 68.9/47 / 56.0/53 & 2.2 / 1.4 & $< 0.9$ / $< 1.5$ & 3.63$^{+0.07}_{-0.23}$ / --- & 12.5$^{+11.0}_{-12.3}$ / $< 40.7$ & 2.4 / 1.6 \\
19 & Triangulum & 56.7/49 / --- & 2.2 / --- & $< 1.4$ / --- & --- / --- & $< 47.1$ / --- & 0.7 / --- \\
\hline
 \end{tabular}
\end{table*}

\begin{figure*}
  \centering
  \includegraphics[width=0.49\linewidth]{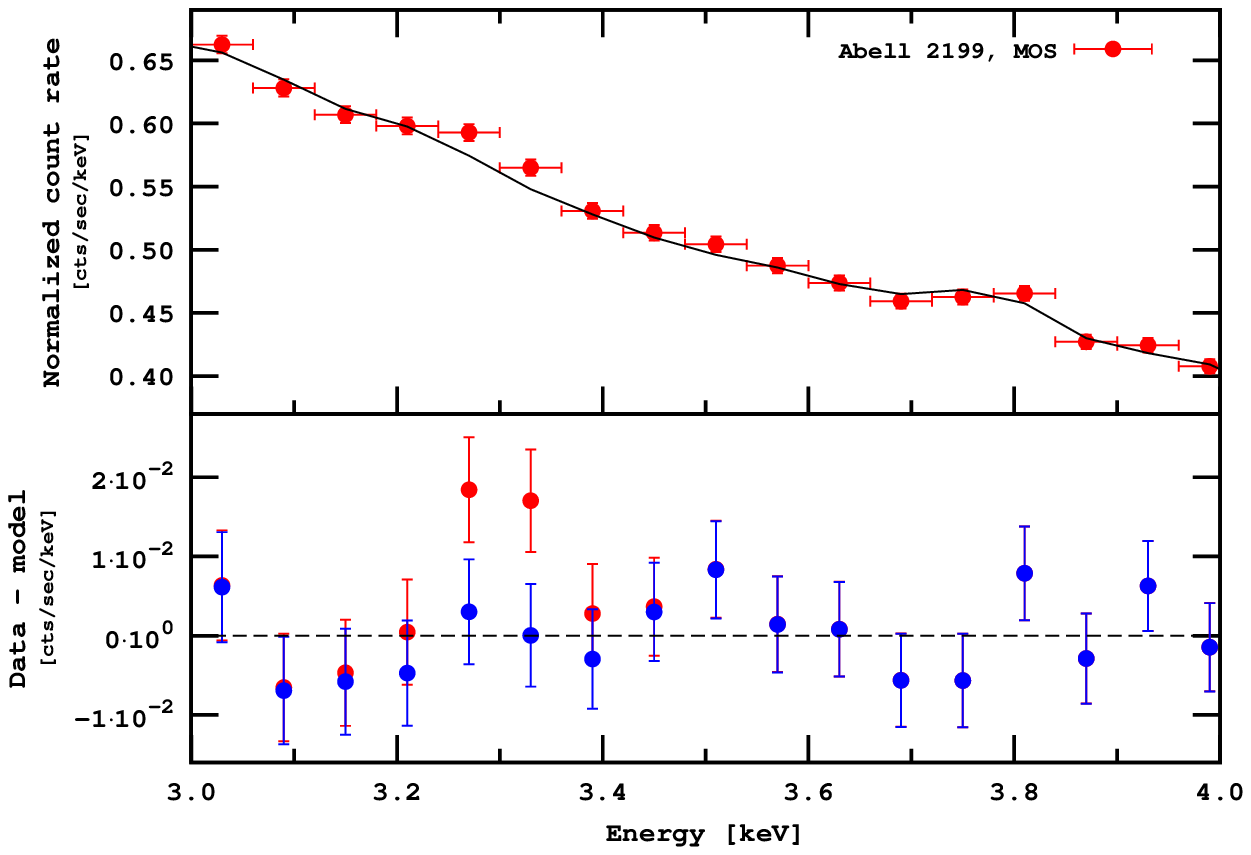}
    \includegraphics[width=0.49\linewidth]{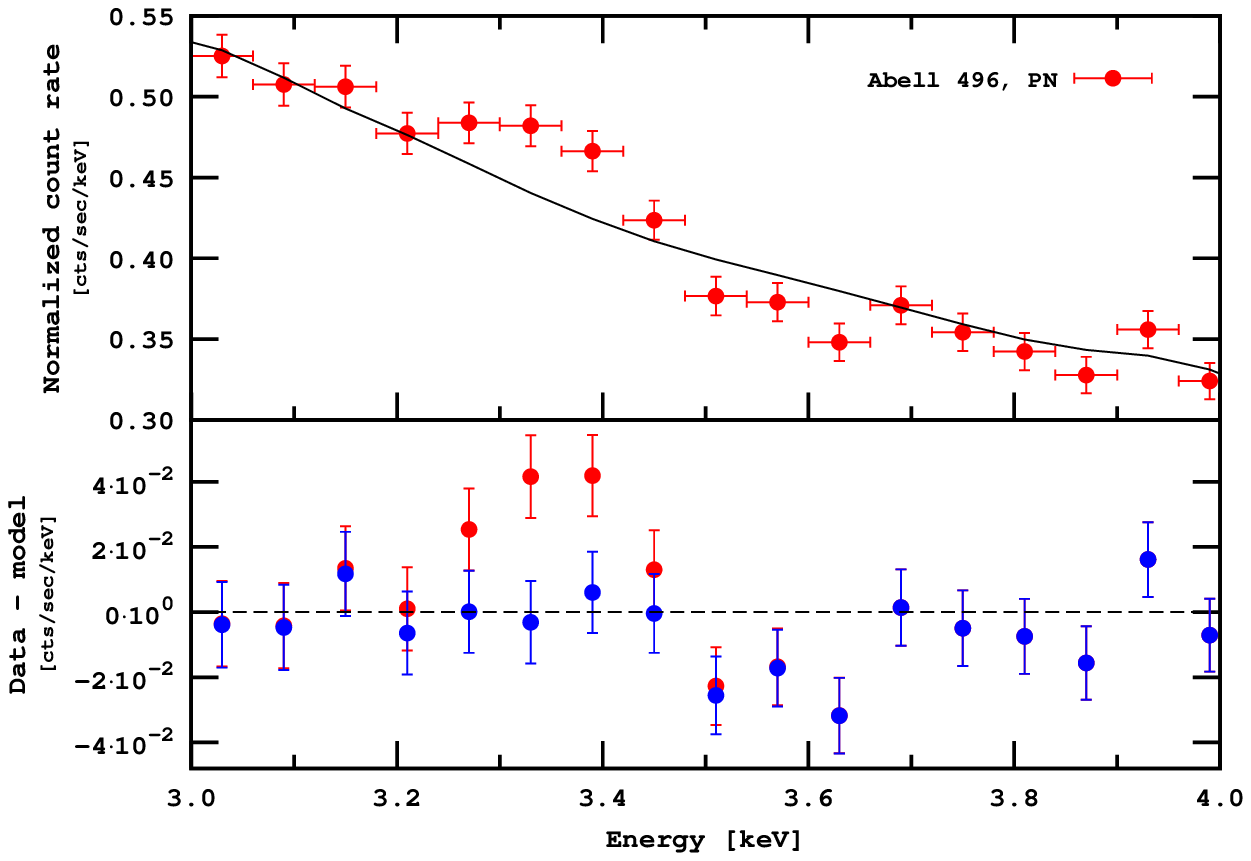}
  \caption{Examples of spectral dataset with identified extra line, see 
  Table~\protect\ref{tab:clusters-model} for details. The spectra are binned by 60~eV and presented in 
  detector's frame similar to~\protect\cite{Boyarsky:14a}.
  Blue and red residuals (bottom) are shown with respect to the best-fit model with and without adding
  an extra line, respectively.
  \textit{Left:} MOS spectrum of Abell~2199. \textit{Right:} PN spectrum of Abell~496.}
  \label{fig:clusters-spectra}
\end{figure*}

\textit{Discussion}.
Assuming decaying dark matter origin of the line at $\sim$3.55~keV previously reported 
by~\cite{Bulbul:14a,Boyarsky:14a,Boyarsky:14b,Urban:14}, we identified 19 galaxy clusters with the largest 
expected significance of dark matter decay signal. 
Using publicly available \xmm\ observations of their central parts, 
we confirmed previous detections in Perseus~\cite{Bulbul:14a,Urban:14} and Coma~\cite{Urban:14} 
clusters, and found $>$2~$\sigma$ positive line-like residuals in 6 new objects, 
see Table~\ref{tab:clusters-model} for details.
We consider the following traditional origins of new line detections:
(a) pure statistical fluctuations; (b) contribution from nearby astrophysical emission lines;
(c) (unknown) systematical effect. 

To check whether pure statistical fluctuations may cause our detections, we simulated cluster spectra 
using FTOOL \texttt{fakeit} for all objects shown in Table~\ref{tab:clusters-model} based on our best-fit 
models without adding a new line in 3.40-3.65~keV, looked for $\Delta\chi^2$ increase caused by 
adding the new narrow \texttt{zgaussian} line in the 
energy range 3.40-3.65~keV (thus accounting for the look-elsewhere effect). 
The average value of 3 maximal $\Delta\chi^2$ for each simulation is in the range 2.1-5.2, 
much smaller than 13.6 (12.5) obtained from our MOS (PN) observations.
Therefore, we conclude that pure statistical fluctuations \emph{alone} can not be responsible for 
line detections in Table~\ref{tab:clusters-model}.

Explanation of the new lines with astrophysical line contribution is also unlikely. The maximal contribution of 
the most promising astrophysical line candidate -- K~XVIII line complex at $\sim$3.51~keV
-- is already included to our model, see Table~\ref{tab:clusters-model}. Other astrophysical lines are both too 
faint and should produce detectable signatures at other energies. For example, to explain the excess in Virgo 
cluster (also consistent with pure statistical fluctuation), 
one should assume 3.398~keV astrophysical line from S~XVI $\sim$5 times higher than the maximal 
contribution from the bright S~XV line complex at $\sim$2.63~keV obtained from Fig.~\ref{fig:S-lines-ratio}.
On the other hand, one cannot exclude the possibility of strongly super-solar abundance because there can be 
variations of Potassium abudance up to 1 dex~\cite{Romano:10,Phillips:15}. According to~\cite{Iakubovskyi:15a},
further studies of the new line using forthcoming observations
of Soft X-ray spectrometer on-board \textit{Astro-H}\, X-ray observatory~\cite{Mitsuda:14} of 
\textit{Micro-X}\, sounding rocket experiment~\cite{Figueroa-Feliciano:15} 
with superior spectral resolution $\lesssim 4-7$~eV can reveal its astrophysical origin. 

\begin{figure}
  \centering
  \includegraphics[width=0.99\linewidth]{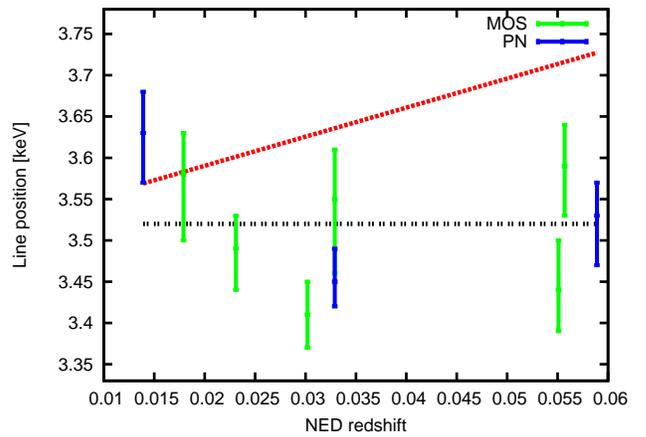}
  \caption{Position of new line detections in cluster's rest frame as function of redshift. 
  Only detections at $>$~2$\sigma$ 
  (corresponding to $\Delta \chi^2 > 6.2$ for 2~d.o.f.) are shown. Red and black dashed lines 
  show expected behavior in case of purely systematic and cosmic line origins (assuming line position 3.52~keV 
  in detector's frame expected from~\protect\cite{Boyarsky:14a,Boyarsky:14b}), respectively.}
  \label{fig:line-position-z}
\end{figure}

The systematic origin of the new line is shown unlikely in pioneering papers~\cite{Bulbul:14a,Boyarsky:14a}. 
In addition, we plotted in Fig.~\ref{fig:line-position-z} the dependence of the line position 
from the object's redshift. 
If the new line were due to systematic effects,
one would expect the corresponding new line in nearby ($z = 0$) objects at $\sim$3.40~keV, 
in apparent tension with observations ($3.53\pm 0.03$~keV for M31~\cite{Boyarsky:14a} and 
$3.539\pm 0.011$~keV for Milky Way~\cite{Boyarsky:14b}) which in turn are better consistent with the new line 
generation in cosmic objects. The mean value of the line positions in 
Fig.~\ref{fig:line-position-z} is 3.52~keV. The average spread between line positions is 75~eV 
close to $\sigma_{\text{instr}} \approx 60$~eV and consistent with our simulations, 
according to that the position of $\sim 3\sigma$ line can be recovered with $\pm$110~eV 
precision in 90\% of cases.

\begin{figure}
  \centering
  \includegraphics[width=0.99\linewidth]{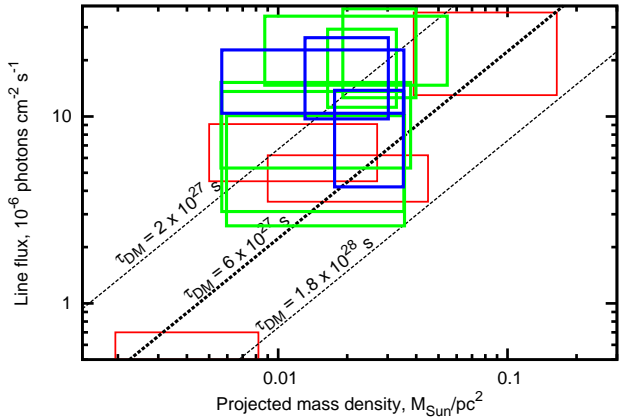}
  \caption{Dependence of the new line flux from the expected projected dark matter mass. This Figure is taken
  from~\protect\cite{Boyarsky:14b}; overplotted are the ranges for our $> 2\sigma$ MOS (green) and PN 
  (magenta) detections. No negative detections are shown here; the strongest restriction
  on $\tau_\dm \gtrsim 3.5\times 10^{27}$~s comes from Virgo cluster, see text.}
  \label{fig:flux-vs-Mproj-blank-v2-Weber-05sigma-newobjects}
\end{figure}

Interpreting the new line due to decaying dark matter (\ref{eq:taudm}) gives the radiative decay lifetime 
$\tau_\dm \approx (3-6)\times 10^{27}$~s
consistent with previous 
detections~\cite{Bulbul:14a,Boyarsky:14a,Boyarsky:14b,Urban:14,Lovell:14}, see 
Fig.~\ref{fig:flux-vs-Mproj-blank-v2-Weber-05sigma-newobjects}. 
Non-detection of the line in some of our galaxy clusters does not exclude the dark matter line origin;
the strongest $2\sigma$ upper bound for our objects comes from Virgo cluster: 
$\tau_\dm \gtrsim 3.5\times 10^{27}$~s. 

Non-detection of $\sim$3.55~keV line in stacked
dSphs by~\cite{Malyshev:14} is also mildly consistent with these results; 
planned observations of Draco dSph would reveal the decaying dark matter nature of the line. 
The absence of the new line in stacked galaxy 
spectra of~\cite{Anderson:14} formally excludes $\tau_\dm < 1.8\times 10^{28}$~s but taking into account 
systematical effects in spectra (e.g. causing significant negative residuals) 
and apparent uncertainty in dark matter distributions~\cite{Boyarsky:09b} produces much weaker bound,
e.g. $\tau_\dm \gtrsim 3.5\times 10^{27}$~s~\cite{Iakubovskyi:14} using stacked dataset of nearby galaxies 
of~\cite{Iakubovskyi:13} with comparable exposure. 
Other bounds on decaying dark matter in $\sim$3.55~keV energy range 
(see~\cite{Iakubovskyi:13,Horiuchi:13,Sekiya:15} and references therein) are also consistent with our detections
after taking into account residual systematic effects and/or uncertainties of dark matter distributions.

\textit{Acknowledgments.} 
We thank Jeroen Franse, Denys Malyshev, Maxim Markevitch and Oleg Ruchayskiy for careful reading of 
the manuscript and their comments.
This work was supported by part by the Swiss National Science Foundation grant SCOPE IZ7370-152581, 
the Program of Cosmic Research of the National Academy of 
Sciences of Ukraine, the State Fund for Fundamental Research of Ukraine 
and the State Programme of Implementation of Grid Technology in Ukraine.

%


\let\jnlstyle=\rm\def\jref#1{{\jnlstyle#1}}\def\aj{\jref{AJ}}
  \def\araa{\jref{ARA\&A}} \def\apj{\jref{ApJ}\ } \def\apjl{\jref{ApJ}\ }
  \def\apjs{\jref{ApJS}} \def\ao{\jref{Appl.~Opt.}} \def\apss{\jref{Ap\&SS}}
  \def\aap{\jref{A\&A}} \def\aapr{\jref{A\&A~Rev.}} \def\aaps{\jref{A\&AS}}
  \def\azh{\jref{AZh}} \def\baas{\jref{BAAS}} \def\jrasc{\jref{JRASC}}
  \def\memras{\jref{MmRAS}} \def\mnras{\jref{MNRAS}\ }
  \def\pra{\jref{Phys.~Rev.~A}\ } \def\prb{\jref{Phys.~Rev.~B}\ }
  \def\prc{\jref{Phys.~Rev.~C}\ } \def\prd{\jref{Phys.~Rev.~D}\ }
  \def\pre{\jref{Phys.~Rev.~E}} \def\prl{\jref{Phys.~Rev.~Lett.}}
  \def\pasp{\jref{PASP}} \def\pasj{\jref{PASJ}} \def\qjras{\jref{QJRAS}}
  \def\skytel{\jref{S\&T}} \def\solphys{\jref{Sol.~Phys.}}
  \def\sovast{\jref{Soviet~Ast.}} \def\ssr{\jref{Space~Sci.~Rev.}}
  \def\zap{\jref{ZAp}} \def\nat{\jref{Nature}\ } \def\iaucirc{\jref{IAU~Circ.}}
  \def\aplett{\jref{Astrophys.~Lett.}}
  \def\apspr{\jref{Astrophys.~Space~Phys.~Res.}}
  \def\bain{\jref{Bull.~Astron.~Inst.~Netherlands}}
  \def\fcp{\jref{Fund.~Cosmic~Phys.}} \def\gca{\jref{Geochim.~Cosmochim.~Acta}}
  \def\grl{\jref{Geophys.~Res.~Lett.}} \def\jcp{\jref{J.~Chem.~Phys.}}
  \def\jgr{\jref{J.~Geophys.~Res.}}
  \def\jqsrt{\jref{J.~Quant.~Spec.~Radiat.~Transf.}}
  \def\memsai{\jref{Mem.~Soc.~Astron.~Italiana}}
  \def\nphysa{\jref{Nucl.~Phys.~A}} \def\physrep{\jref{Phys.~Rep.}}
  \def\physscr{\jref{Phys.~Scr}} \def\planss{\jref{Planet.~Space~Sci.}}
  \def\procspie{\jref{Proc.~SPIE}} \let\astap=\aap \let\apjlett=\apjl
  \let\apjsupp=\apjs \let\applopt=\ao \def\jcap{\jref{JCAP}}

\appendix
\onecolumngrid

\section{Dark matter distributions in galaxy clusters}
\label{app:dm-distributions}

To describe dark matter distribution in galaxy clusters used in our work 
we compliled in Table~\ref{tab:cluster-profiles} dark matter distributions from the literature using the 
extended dataset of~\cite{Boyarsky:09b}.

All cluster distributions are described with
Navarro-Frenk-White (NFW) profile~\cite{Navarro:96}
\begin{equation}
  \rho_\nfw(r) = \frac{\rho_s r_s}{r(1+r/r_s)^2}\label{eq:rho_NFW}
\end{equation}
parametrised by $\rho_s$ and $r_s$.

The dark matter column density inside \xmm\ field-of-view radius 
$R_{14} = D_L \times \frac{14\pi}{60\times 180}$ is derived as
\begin{equation}
 \label{eq:Sbar}
 \S = \frac2{R_{14}^2}\int^{R_{14}}_0 rdr \int  dz\, \rho_\nfw(\sqrt{r^2+z^2})
\end{equation}

For the NFW density distribution~(\ref{eq:rho_NFW}):
\begin{equation}{\S}_\nfw(R) = \frac{4\rho_s
    r_s^3}{R^2}\left[\frac{\arctan\sqrt{R^2/r_s^2-1}}{\sqrt{R^2/r_s^2-1}} +
    \log\left(\frac{R}{2r_s}\right)\right]\label{eq:S_NFW}\;.
\end{equation}

Dark matter distribution parameters for our Galaxy are taken from~\cite{Boyarsky:14b}.

\begin{table*} [!tb]
  \centering 
\begin{tabular}[c]{lccccccccccccccccc}
  \hline
 Object & Reference & Profile & $R_{14}$ & $r_s$ & $\rho_s$ & $\S_{obj}$ \\
        &         &         & kpc   & kpc & $10^6 \text{M}_\odot/\text{kpc}^3$ & M$_\odot / \text{pc}^2$ \\
  \hline
Virgo & \cite{McLaughlin:98} & NFW & 73 & 560 & 0.32 & 808 \\
\hline
Centaurus & \cite{Ettori:02} & NFW & 180 & 345 & 1.51 & 1087 \\
\hline
Abell~85 & \cite{Ettori:02} & NFW & 867 & 1282 & 0.25 & 549 \\
Abell~85 & \cite{Wojtak:10} & NFW & 922 & 650 & 0.37 & 210 \\
\hline
Abell~478 & \cite{Mahdavi:07} & NFW & 1978 & 1140 & 0.85 & 686 \\
Abell~478 & \cite{Pointecouteau:05} & NFW & 1518 & 488 & 0.77 & 134 \\
\hline
Abell~2199 & \cite{Ettori:02} & NFW & 526 & 560 & 0.76 & 552 \\
Abell~2199 & \cite{Rines:03} & NFW & 509 & 214 & 1.7 & 180 \\
\hline
Abell~496 & \cite{Ettori:02} & NFW & 545 & 738 & 0.45 & 530 \\
Abell~496 & \cite{Wojtak:10} & NFW & 550 & 420 & 0.48 & 191 \\ 
\hline
2A0335+096 & \cite{Ettori:02} & NFW & 593 & 626 & 0.52 & 420 \\
2A0335+096 & \cite{Voigt:06} & NFW & 593 & 130 & 3.6 & 101 \\
\hline
Abell~1060 & \cite{Richtler:11} & NFW & 195 & 140 & 7.2 & 899 \\
Abell~1060 & \cite{Wojtak:10} & NFW & 211 & 140 & 5.8 & 667 \\
\hline
Abell~3266 & \cite{Ettori:02} & NFW & 991 & 1576 & 0.19 & 543 \\
\hline
Abell~S805 & \cite{Wojtak:10} & NFW & 232 & 190 & 2.0 & 386 \\ 
\hline
Coma & \cite{Ettori:02} & NFW & 397 & 459 & 1.23 & 788 \\
Coma & \cite{Gavazzi:09} & NFW & 407 & 326 & 0.85 & 275 \\
\hline
Abell~S239 & \cite{Democles:10} & NFW & 1095 & 792 & 0.55 & 391 \\
Abell~S239 & \cite{Democles:10} & NFW & 1095 & 576 & 0.98 & 361 \\
\hline
Abell~2142 & \cite{Ettori:02} & NFW & 1469 & 1654 & 0.18 & 406 \\
Abell~2142 & \cite{Wojtak:10} & NFW & 1520 & 990 & 0.18 & 144 \\
\hline
Abell~2319 & \cite{Ettori:02} & NFW & 943 & 1301 & 0.24 & 506 \\ 
\hline
Abell~1795 & \cite{Ettori:02} & NFW & 1052 & 1024 & 0.34 & 417 \\
Abell~1795 & \cite{Ikebe:04} & NFW & 1052 & 393 & 0.82 & 139 \\
\hline
Abell~209 & \cite{Okabe:10} & NFW & 3460 & 2513 & 0.18 & 408 \\
Abell~209 & \cite{Bardeau:07} & NFW & 3435 & 502 & 0.386 & 24 \\
\hline
Perseus & \cite{Simionescu:11} & NFW & 305 & 360 & 1.1 & 563 \\
\hline
PKS~0745-191 & \cite{Ettori:02} & NFW & 1665 & 1148 & 0.33 & 324 \\
PKS~0745-191 & \cite{George:08} & NFW & 1779 & 230 & 2.5 & 59 \\
\hline
Triangulum & \cite{Ettori:02} & NFW & 856 & 666 & 0.83 & 534 \\
\hline
\end{tabular} 
\caption{Parameters of dark matter distributions of galaxy clusters used in this work.
}
\label{tab:cluster-profiles} 
\end{table*}

\section{Modeling astrophysical lines}
\label{app:lines-modeling}

\begin{figure}[tbh!]
  \centering
  \includegraphics[width=0.99\linewidth]{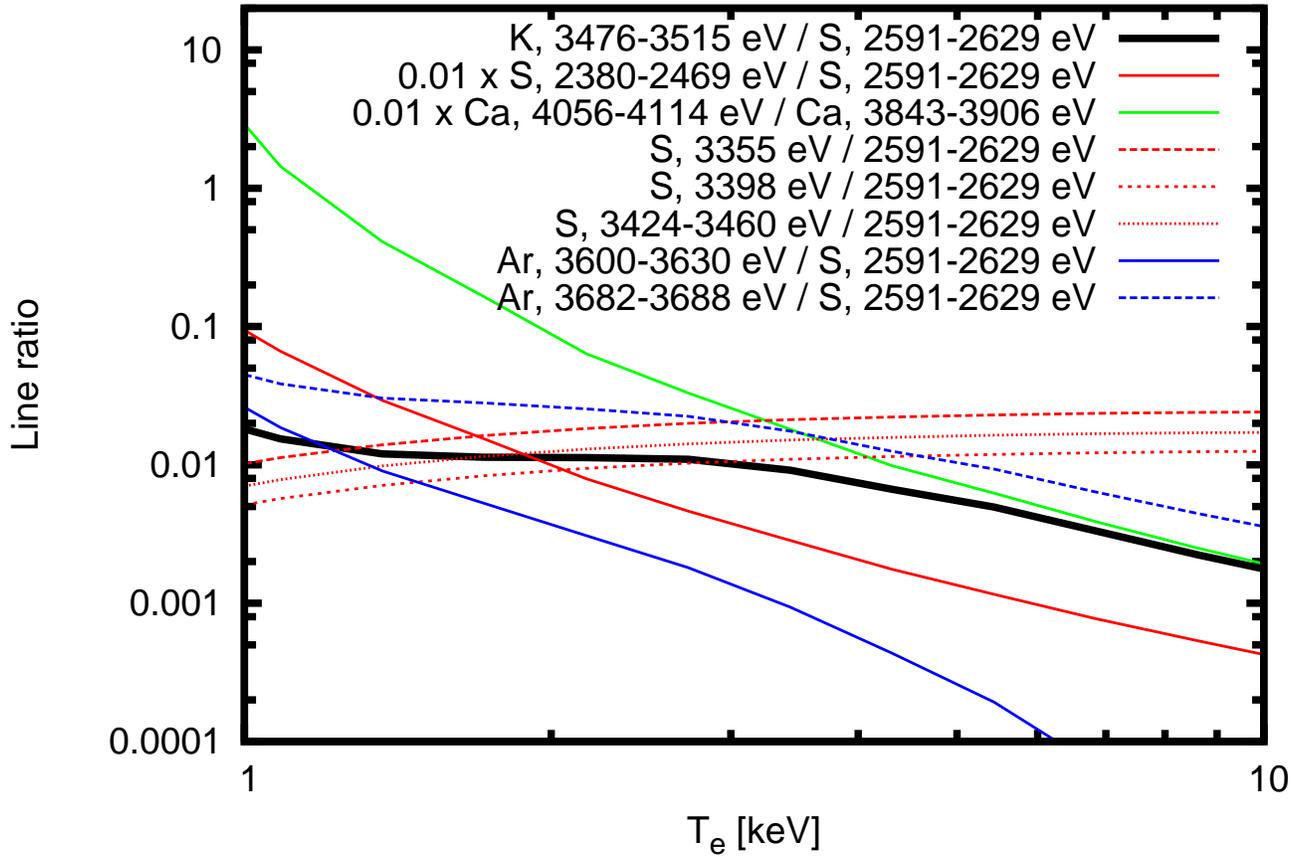}~%
  \caption{Line emissivity ratios for K~XVIII line complex at $\sim$3.51~keV and other 
  emission lines of our interest 
  as functions of the electron temperature $T_e$ in the plasma. The line emissivities are calculated using
  AtomDB version 3.0.3 with line emissivities $> 10^{-22}~\unit{ph\ cm^3/s}$.}
  \label{fig:S-lines-ratio}
\end{figure}

To check the astrophysical origin of the new line, we added narrow \texttt{zgaussians} corresponding to
known astrophysical lines in this range. For example, according to the newest atomic database AtomDB v.~3.0.3,
there are S~XVI line complexes at 3.355, 3.398, 3.424, 3.441, 3.452 and 3.460~keV. Fortunately, the intensity
of these lines can be robustly predicted by the measured S~XV line complex at $\sim$2.63~keV,
see Fig.~\ref{fig:S-lines-ratio} for details. 

We paid special attention to potential contribution from K~XVIII lines near 3.51~keV,
see e.g.~\cite{Iakubovskyi:15a} for details. The distance between these lines is 
smaller than the energy resolution of \xmm/EPIC, so we modeled 
the K~XVIII line complex as a single \texttt{zgaussian}
with mean energy 3.51~keV. Because there is no ``reference'' Potassium line 
to reproduce the 3.51~keV line flux, we fixed only the upper bound of the 3.51~keV line intensity relating it
to S~XVI line flux at 2.63~keV (or, if 2.63~keV is not detected in the dataset -- to its 2$\sigma$ upper bound)
using the procedure described in~\cite{Iakubovskyi:15a}. 
To derive electron temperature, we used flux ratios of strong elemental lines, namely S~XV lines at 2.45~keV,
S~XVI lines at 2.63~keV, Ca~XIX lines at 3.90~keV and Ca~XX lines at 4.10~keV.
Because 3.51/2.63~keV line ratio is a 
decreasing function of electron temperature $T_e$, see Fig.~\ref{fig:S-lines-ratio},
we used minimal temperature $T_{e,line} = \text{min}\left[T_{e,S},T_{e,Ca}\right]$ for conservative estimate.
Another source of uncertainty 
comes from the (largely unknown) relative K/S abundance ratio. To account possible 
uncertainties~\cite{Romano:10,Jeltema:14a,Phillips:15} 
we allowed this ratio to be up to 3~Solar values of~\citep{Anders:89}. Note that from comparison of 
columns~4 and~6 one can derive that to explain new line emission solely in terms of K~XVIII line complex at 
3.51~keV, one should assume strongly supersolar (Abund[K]/Abund[S] $> $15~Solar) ratios for \emph{all} 
our of detections.

\end{document}